\documentclass[emulateapj]{emulateapj}

\slugcomment{Draft version November 22, 2006}
\shorttitle{Inclination-Dependent Luminosity Function of Spiral
Galaxies} \shortauthors{ Z. Shao, Q. Xiao, S. Shen, others}
\received{} \accepted{}

\begin{document} \title{Inclination-Dependent Luminosity Function of
Spiral Galaxies in the Sloan Digital Sky Survey: Implication for
Dust Extinction}

\author{ Zhengyi Shao\altaffilmark{1,2}, Quanbao Xiao\altaffilmark{1,2},
Shiyin Shen\altaffilmark{1,2}, H. J. Mo\altaffilmark{3}, Xiaoyang
Xia\altaffilmark{4}, Zugan Deng\altaffilmark{5}}
\altaffiltext{1}{Shanghai Astronomical Observatory, CAS, Shanghai
200030, P.R. China} \altaffiltext{2}{Joint institute for galaxy \&
 cosmology, CAS, Shanghai 200030,  P.R.
China}\altaffiltext{3}{Astronomy Department, University of
Massachusetts, Amherst MA 01003, USA} \altaffiltext{4}{Department
of Physics, Tianjin Normal University, Tianjin 300074, P. R. China
} \altaffiltext{5}{Graduate School, Chinese Academy of Sciences,
Beijing 100080, P. R. China}

\email{zyshao@shao.ac.cn}

\begin{abstract}
Using a samples of 61506 spiral galaxies selected from the SDSS
DR2, we examine the luminosity function (LF) of spiral galaxies
with different inclination angles. We find that the characteristic
luminosity of the LF, $L^*$, decreases with increasing
inclination, while the faint-end slope, $\alpha$, depends only
weakly on it. The inclination-dependence of the LF is consistent
with that expected from a simple model where the optical depth is
proportional to the cosine of the inclination angle, and we use a
likelihood  method to recover both the coefficient in front of the
cosine, $\gamma$, and the LF for galaxies viewed face-on. The
value of $\gamma$ is quite independent of galaxy luminosity in a
given band, and the values of $\gamma$ obtained in this way for
the 5 SDSS bands give an extinction curve which is a power law of
wavelength ($\tau\propto\lambda^{-n}$), with a power index
$n=0.96\pm0.04$. Using the dust extinction for  galaxies obtained
by Kauffmann et al. (2003), we derive an `extinction-corrected'
luminosity function for spiral galaxies. Dust extinction makes
$M^*$ dimmer by about $0.5$ magnitudes in the $z$-band, and about
$1.2$ magnitudes in the $u$- band. Since our analysis is based on
a sample where selection effects are well under control, the
dimming of edge-on galaxies relative to face-on galaxies is best
explained by assuming that galaxy disks are optically thick in
dust absorptions.
\end{abstract}

\keywords{galaxies: spiral --- galaxies: luminosity function ---
ISM: dust, extinction}

\section{Introduction}

Dust plays an important role in the observed properties of
galaxies. It not only causes galaxies to dim but also causes them
to redden. In order to obtain the intrinsic properties of galaxies
from observation, it is necessary to understand dust extinction.
Dust extinction is believed to be more important in late-type
spiral galaxies than in early-type galaxies, since spiral galaxies
are richer in cold gas and have on-going star formation (e.g.
Calzetti 2001 for a review and references therein). Dust opacity
in spiral disks can be probed by studying the average photometric
properties of spiral galaxies as a function of disk inclination
angle, as first suggested by Holmberg (1958, 1975). Theoretically,
it can be shown that, for completely opaque disks, the surface
brightness to be observed is almost independent of inclination
angle, while the luminosity dims as the inclination angle changes
from face-on to edge-on. On the other hand, for completely
transparent disks, the luminosity is independent of inclination
while the surface brightness brightens as the disk changes from
face-on to edge-on. However, the power of the Holmberg test is
limited by the finite sizes of galaxy samples, because it is in
general difficult to accurately estimate the inclination angles
for individual galaxies. The axis ratios of galaxy images are
usually used to estimate the inclination angles. For an
infinitesimally thin and round disk, the inclination angle
$\theta$ can be obtained directly from the axis ratio $b/a$
through $\cos\theta=b/a$. Unfortunately, real disks are neither
completely round nor infinitesimally thin, and the observed axis
ratio depends not only on the inclination angle but also on the
ellipticity and thickness of the galaxy. Unless such dependence is
fully taken into account, the results from the Holmberg test
cannot be interpreted straightforwardly. In addition, the test
based on the average photometric properties of galaxies as
functions of inclination angle may be further complicated by the
incompleteness and selection bias of galaxy sample. For example,
if disk galaxies are optically thin, a galaxy sample may be biased
for low-luminosity galaxies with high inclinations because of the
the enhanced surface brightness. With such selection bias, face-on
galaxies will on average be brighter than edge-on galaxies, which
may be falsely interpreted as the disks being optically thick.
Because of these limitations, the conclusion about the opacity of
the disk galaxies are still controversial (e.g. Burstein Haynes \&
Faber 1991; Byun 1993; Giovanelli et al. 1994, 1995; Xilouris et
al. 1999; Masters, Giovanelli \& Haynes 2003; Holwerda et al.
2005).

 With a large and well-defined sample, the above-mentioned difficulties
in the Holmberg test can in principle be overcome. As shown in
Ryden (2004), a large sample allows one to estimate the thickness
and ellipticity  robustly from the distribution of axis ratios in
a statistical way. Moreover, with a large, well-defined sample,
one can study the conditional distribution functions of
photometric properties for given inclination angles, instead of
just considering the mean photometric properties of galaxies.
Thus, the selection bias and sample incompleteness can be taken
into account strictly in a statistical sense. With the advent of
the Sloan Digital Sky Survey
(SDSS, York et al. 2000), it is now possible to make such
analysis.

In this paper, we use the SDSS galaxy sample to study dust
extinctions in spiral galaxies. Our analysis is based on the
luminosity function (hereafter LF) of galaxies as a function of
inclination angles. We statistically correct the axis ratios of
galaxy images to obtain their inclination angles based on the
method of Ryden (2004). We compare the change of LF with
inclination angle in different wavebands to constrain the shape of
the dust extinction curve. The paper is organized as follows.  The
selection of sample is described in \S~\ref{sec:sample} and
quantities to specify the inclinations of galaxies are discussed
in \S~\ref{sec:inclination}. In \S~\ref{sec:LF}, after a brief
description of the LF estimators adopted in this paper
(\S~\ref{sec:estimator}), we present the results of the LFs in all
the 5 SDSS bands for galaxies of different inclination angles
(\S~\ref{sec:result}). In \S~\ref{sec:IDE}, the
inclination-dependence of the LF is modelled in terms of dust
extinction. The possible luminosity dependence of the dust
extinction and the dust-corrected LF of spiral galaxies are
discussed in \S~\ref{sec:discussion}. Finally, our results are
summarized in \S~\ref{sec:summary}.

\section{Observational data}\label{sec:data}

\subsection{Galaxy samples}\label{sec:sample}

The galaxy samples we used were selected from  the New York
University Value-Added Galaxy Catalog (NYU-VAGC, Blanton et al.,
2005) of the SDSS second data release (DR2, Abazajian et al.
2004). The redshift catalog of the DR2 covers 2,627 deg$^2$ of the
celestial sphere and photometric data in the 5 SDSS wave bands,
$u$, $g$, $r$, $i$ and $z$, are available for each of the galaxies
(Abazajian et al. 2004) directly from the SDSS pipeline. The
NYU-VAGC includes additional information for extragalactic targets,
such as $K$-correction, spectroscopic target completeness, etc..
From the NYU-VAGC, we select spiral galaxies with $fracdeV_r \leq
0.5$, where the photometric parameter $fracdeV$ is a point spread
function (PSF) corrected indicator of galaxy morphology. In the
SDSS pipeline, each galaxy was fitted by an exponential profile
and a de Vaucouleurs' profile. The best linear combination of
these two profiles was used to represent the profile of the
galaxy, and $fracdeV$ is the fraction of luminosity contributed by
the de Vaucouleurs' profile. Bernardi et al. (2005) used
$fracdeV_r \geq 0.8$ to select early-type galaxies. We use
$fracdeV_r \leq 0.5$ to ensure that the galaxies we select are
dominated by the exponential component.

To construct samples to study the LFs of galaxies, both redshift
and flux limits are applied. Only galaxies with redshift in the
range $0.01 \leq z_{\rm red} \leq 0.22$ are selected. The upper
limit is imposed to minimize the uncertainty in the $K$-correction
and possible redshift evolutions of galaxies; the lower limit is
employed to avoid the nearest galaxies with large uncertainties of
distance caused by peculiar velocities. A flux limit in $r$ band,
$m_r=17.60$, is chosen to ensure the completeness in spectroscopy.
For other wave bands, the flux limits are chosen to be
sufficiently high, so that almost all of the galaxies ($99\%$)
brighter than the limit in the band in consideration have $r$-band
flux brighter than $17.60$ mag. The flux limits of all 5 bands we
use and the numbers of galaxies in the corresponding samples are
listed in Table~\ref{tab:limits}. Note that we use the Petrosian
(1976) magnitude to refer the luminosities of galaxies. This
definition of magnitude has the advantage that it is a uniform
measurement of flux quite independent of the distance of the
target. Furthermore, for an exponential profile the Petrosian
magnitude contains almost all of the flux of the source (e.g.
Stoughton et al. 2002).


\begin{table}
\caption{The Sample of galaxies in Each SDSS Band}
\begin{tabular}{cccrr} \hline \hline
Band & Flux Limit & Redshift Limit & $N_{\rm gal}$ & $N_{\rm spiral}$\\
\tableline
 $u$  & $14.50 < m < 18.60$ & $0.01 < z < 0.22$ &  42033 & 24106 \\
 $g$  & $14.50 < m < 17.90$ & $0.01 < z < 0.22$ &  93221 & 43252 \\
 $r$  & $14.50 < m < 17.60$ & $0.01 < z < 0.22$ & 162279 & 61506 \\
 $i$  & $14.50 < m < 17.15$ & $0.01 < z < 0.22$ & 148934 & 52336 \\
 $z$  & $14.50 < m < 16.85$ & $0.01 < z < 0.22$ & 136151 & 43653 \\
\tableline \label{tab:limits}
\end{tabular}
\tablenotetext{}{$N_{\rm gal}$ is the number of galaxies selected
within the flux and redshift limits. $N_{\rm spiral}$ is the
number of spiral galaxies selected with $fracdeV_r \leq
0.5$.\vspace{8mm}}

\end{table}

For each band, the absolute magnitude $M$  is calculated using
\begin{equation}\label{eq:Mabs}
   M_{\lambda} = m_{\lambda} - 5\log(d_L/10pc)-K_{\lambda}(z)
\end{equation}
where $\lambda$ is the wave band in consideration,
$m_\lambda$ is the apparent magnitude corrected for Galactic
extinction based on the $E(B-V)$ maps of Schlegel, Finkbeiner \&
Davis (1998), $d_L$ is the luminosity distance, and $K_\lambda(z)$ is
the $K$-correction value taken from NYU-VAGC. Throughout the
paper, we use the standard $\Lambda-$cosmology with
$\Omega_0=0.3$, $\Omega_\Lambda=0.7$ and $H_0=100
h^{-1}\rm{kms^{-1}Mpc}$.

\subsection{Inclination parameters}\label{sec:inclination}

 The apparent axis ratio of the image of a galaxy, which is provided
by the SDSS pipeline, is a direct measurement of the inclination
of spiral galaxies. If disks are thin and round, the
minor-to-major axis ratio $b/a$ is related to the inclination
angle $\theta$ (defined to be the angle between the line-of-sight
and the axis of angular momentum of the disk) by $b/a=\cos\theta$.
In this paper, we use the $r$ band axis ratio $ab\_exp_r$, taken
from the best fit of the images of galaxies with an exponential
profile convolved with the PSF (see Stoughton et al. 2002). Since
the sample galaxies we selected are expected to be dominated by
the exponential component ($fracdeV_r \leq 0.5$), the parameter
$ab\_exp_r$ (hereafter referred to as $b/a$) should be a
reasonable representation of the apparent axis ratio.

\begin{figure}
  \includegraphics[height=0.4\textwidth,angle=-90]{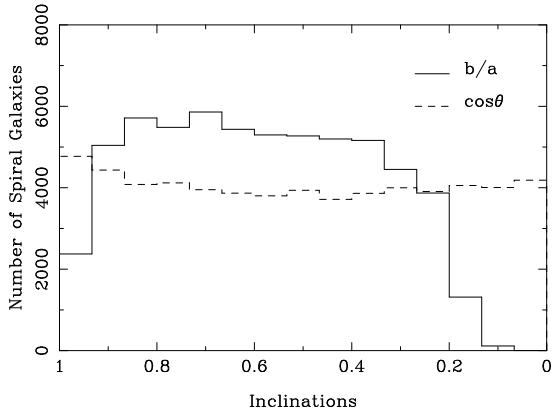}\\
  \caption{The distributions of the apparent axis ratio $b/a$ (solid line)
  and of the simulated inclinations $\cos \theta$ (dashed line).}
 \label{fig:b/a}\vspace{8mm}
\end{figure}

Figure~\ref{fig:b/a} shows the distribution of $b/a$ for the
sample galaxies (solid line). If disks are completely round and
infinitesimally thin, and if their rotation axes are randomly
orientated in space, the distribution of $b/a$ should be flat. For
our sample, the distribution is almost flat for $0.3<b/a<0.9$, but
disks with axis ratios close to $1$ or close to 0 are less common.
The lack of very round images ($b/a \sim 1$) is caused by
the intrinsic ellipticity, $\epsilon$, of spiral galaxies, as
demonstrated in detail by Ryden (2004). The lack of images with
small $b/a$ is caused by two factors. The first is the intrinsic
thickness of galaxy disks: the typical scale height of a disk is
about $10\%$ of the disk scale length, as shown in Giovanelli et
al. (1994). The other is the existence of a central bulge which,
being spheroidal, can thicken the image of a spiral galaxy,
especially an edge-on spiral. In what follows, we do not
distinguish these two different causes of disk thickening. Instead
we use a single `effective thickness' parameter, $\nu$,  defined
as the ratio of scale height to scale length, to represent the
thickness of a spiral galaxy.

 Using a sample of SDSS spiral galaxies that have relative large
apparent size, $R_e > 5\sqrt{T_{\rm psf}}$,  where $R_e$ is the
effective radius of a exponential profile and $T_{\rm psf}$ is the
adaptive second moment of the PSF image (so that $\sqrt{T_{psf}}$
may be used as a characteristic scale of the PSF), Ryden (2004)
found that both $\nu$ and $\log\epsilon$ have roughly Gaussian
distributions. For example, in the $r$-band, Ryden (2004) found
that the median ($\nu_0$) and dispersion ($\sigma_\nu$)
of the $\nu$-distribution are $0.216$ and $0.056$, respectively,
while those for the $\log\epsilon$-distribution are
$\log\epsilon_0=-1.83$ and $\sigma_{\log\epsilon}= 0.93$.
However, with a typical PSF of 1.5 arcsec, the majority of SDSS
galaxies have $R_e < 5\sqrt{T_{\rm psf}}$ (e.g. Shen et al. 2003).
Moreover, the apparent thickness is expected to vary with the
apparent size ($R_e$), with smaller images appearing rounder, even
though the images are de-convolved with the PSF. Galaxies with
small apparent sizes include not only intrinsically small galaxies
at small distances, but also intrinsically luminous galaxies at
large distances. Since the average bulge-to-disk ratio of spiral
galaxies increases with galaxy luminosity, the bulge components
may become more dominant in a sample that contains more distant
galaxies with small apparent sizes. Furthermore the bulge
components are in general easier to observe than the disk
components, because of their higher surface brightness. Both
factors will cause the distant, apparently small spirals to be
rounder. For this reason, we divide our sample into subsamples
according the apparent sizes ($R_e$) of galaxies, and study the
thickness parameter separately for each of the subsamples.

\begin{figure}
  \includegraphics[height=0.4\textwidth,angle=-90]{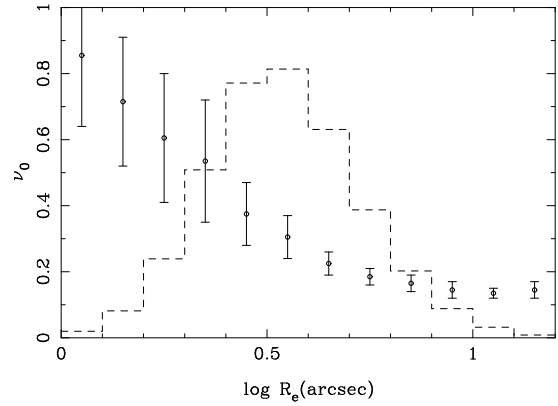}\\
  \caption{The best-fit values of the effective thickness
  parameters $\nu_0$ and $\sigma_\nu$ (shown as error bars) for
  spiral  galaxies with different apparent
  size $\log R_e$ in the $r$-band. Dashed line is the
  distribution of $\log R_e$ for sample galaxies.}
  \label{fig:nu0}\vspace{8mm}
\end{figure}

  We assume that both the thickness parameter $\nu$ and
ellipticity $\log\epsilon$ follow Gaussian distributions and
estimate the thickness parameters $\nu_0$ and $\sigma_\nu$ for
spirals with different apparent sizes using the following
Monte-carlo procedure. For a given galaxy with given shape
parameters ($\nu$ and $\epsilon$) and viewing angles ($\theta ,
\varphi$), the apparent axis ratio $b/a$ can be calculated using
\begin{equation}\label{eq:b/a}
 b/a =
  \left[\frac{A+C-\sqrt{(A-C)^2+B}}{A+C+\sqrt{(A-C)^2+B}}\right]^{1/2}\,,
\end{equation}
where
\begin{equation}\label{eq:A}
    A=[1-\epsilon(2-\epsilon)\sin^2\varphi]\cos^2\theta +
    \nu^2\sin^2\theta ,
\end{equation}
\begin{equation}\label{eq:B}
    B=4\epsilon^2(2-\epsilon)^2 \cos^2\theta \sin^2\varphi
    \cos^2\varphi ,
\end{equation}
\begin{equation}\label{eq:C}
    C=1-\epsilon(2-\epsilon)\cos^2\varphi
\end{equation}
(see Binney 1985). Model galaxies are assumed to have random
distribution in the view angle, and to have random distributions
in $\nu$ and $\log\epsilon$ according to their Gaussian
distribution functions. To reduce model parameters, we assume that
$\log{\overline\epsilon}$ and $\sigma_{\log\epsilon}$ have the
values given by Ryden (2004). The best values of $\overline{\nu}$
and $\sigma_\nu$ are obtained by matching the predicted $b/a$
distribution with the observed $b/a$ distribution (for each
$R_e$-subsample), with the use of the least-square criterion. The
fitting results for $\nu_0$ and $\sigma_\nu$ are plotted in
Figure~\ref{fig:nu0}. As expected, disk galaxies with large
apparent sizes have smaller `effective thickness'. At apparent
size $R_e>8$ arcsec, $\nu_0$ reaches a constant value $\sim 0.14$.
This value may be considered as the upper limit of the effective
thickness of the disks, or be explained as the  intrinsic
thickness of pure disk should be less then 0.14. Since the typical
seeing condition in SDSS has $T_{\rm psf}=4 \rm{Pixel}^2$, the
galaxies selected by Ryden (2004) have $R_e > 5\sqrt{T_{\rm psf}}$
and so are dominated by galaxies with $R_e \ga 4$ arcsec (see
Figure ~\ref{fig:nu0}). Our results for galaxies with $R_e \ga 4$
arcsec are $\nu_0\pm\sigma_\nu \sim 0.195 \pm 0.035$, close to the
results of Ryden $0.216 \pm 0.056$.

\begin{figure}
  \includegraphics[height=0.4\textwidth,angle=0]{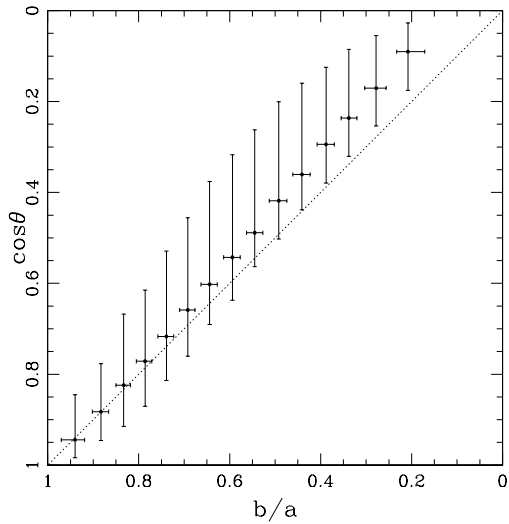}\\
  \caption{The correlation between the apparent axis ratio $b/a$
  and the simulated inclination $\cos \theta$. Points are located
  at the median value of $b/a$ and $\cos\theta$, and error bars
   represent the 16-84 percentiles of the distribution.}
  \label{fig:cos-theta}\vspace{8mm}
\end{figure}

The above procedure also provides a statistical way to relate the
real inclination of a galaxy (represented by $\cos\theta$) to the
apparent axis ratio ($b/a$). To do this, each galaxy is assigned a
possible viewing angle using a Monte Carlo method. For a galaxy
with given (observed) $b/a$ and $R_e$, we first randomly select a
set of shape parameters ($\nu$ and $\epsilon$) from their Gaussian
distributions obtained above, and then calculate the apparent axis
ratio using equation (\ref{eq:b/a}). The value of $\theta$ that
best reproduces the observed $b/a$ is taken to be the viewing angle of
this galaxy. Since our samples are sufficiently large, the
simulated values of $\theta$ should represent the real
distribution of the inclination angles in a statistical sense.

Figure~\ref{fig:cos-theta} shows the correlation between the
simulated $\cos\theta$ and the observed $b/a$. The whole sample is
divided into 15 subsets in $b/a$, with each subset containing a
similar number of galaxies. In the figure, points are median
values while errorbars are the 16-84 percentiles of the
distribution. The distribution of $\cos\theta$ is also plotted in
Figure~\ref{fig:b/a} as the dashed lines. As one can see,
$\cos\theta$ follows a nearly flat distribution as expected.
However, it should be emphasized that $\cos\theta$ is an
inclination parameter that has only statistical meaning. For
individual galaxies, the uncertainty in $\cos\theta$ is as large
as the scatter shown in Figure~\ref{fig:cos-theta}.

\section{Galaxy luminosity function: dependence on inclination}
\label{sec:LF}

\subsection{Luminosity function estimator}
\label{sec:estimator}

 In estimating the luminosity function (LF) of galaxies, we use the
maximum-likelihood method proposed by Sandage, Tammann \& Yahil
(1979, hereafter STY). This method assumes that the LF of galaxies
has the Schechter form (Schechter 1976),
\begin{equation}\label{eq:lf}
\phi(L){\rm d} L=\phi^{*}(L/L^*)^{\alpha}\exp(-L/L^*) {\rm
d}(L/L^*)\,,
\end{equation}
where $L^{*}$ is the characteristic luminosity, $\alpha$ the
faint-end slope, and $\phi^{*}$ the overall amplitude of the LF.
The STY method uses such a luminosity function to calculate the
probability for a galaxy with luminosity $L_i$ and redshift $z_i$ to be
included in a magnitude-limited sample:
\begin{equation}\label{LuminosityF_pi}
p_i=\frac{\phi(L_i) {\rm d} L}{\int_{L_{\rm min}(z_i)}^{L_{\rm
max}(z_i)}\phi(L)\, {\rm d} L}\,,
\end{equation}
where $L_{\rm min}(z_i)$ and $L_{\rm max}(z_i)$ are the lowest and
highest luminosities that a galaxy at redshift $z_i$ can have in
order for it to be included in the sample. The likelihood function
is defined as
\begin{equation}\label{Lum_likelihood}
{\cal L} =\prod_i p_i\,
\end{equation}
where the product extends over all galaxies in the sample, and its
maximization provides an estimate for $\alpha$ and $L^*$ (or
equivalently the corresponding absolute magnitude $M^*$).

\begin{figure}
  \includegraphics[height=0.4\textwidth,angle=0]{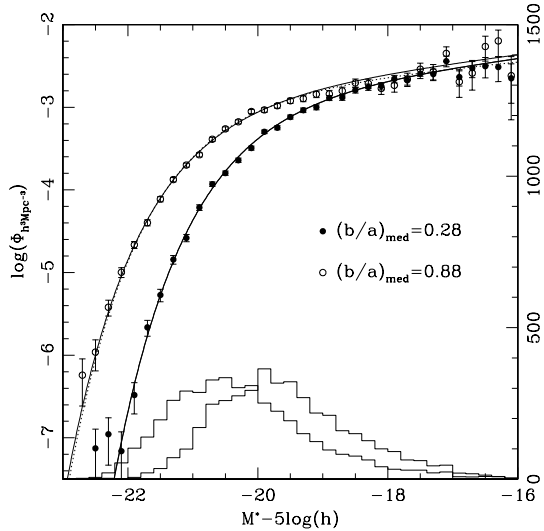}\\
  \caption{The luminosity functions for the nearly face-on
  subsample (open circles) and the nearly edge-on subsample
  (solid circles). Smooth curves are the luminosity functions
  obtained using the STY method. Dotted curves
  show the best fit of the data points with a Schechter function.
  Solid curves are the fit results assuming $\alpha= \langle\alpha_{\rm sub}\rangle
   = -1.25$. Points and errors are obtained from the SWML method.
  Histograms near the bottom are the numbers of galaxies
  in bins of absolute magnitude. The results for
  edge-on galaxies are shifted vertically by a factor of
  two for easy to compare the shape of  LFs.
}
  \label{fig:lf}\vspace{8mm}
\end{figure}

We estimate the LFs for spiral galaxies with different
inclinations in all the 5 SDSS bands. In order to make meaningful
comparisons among the results, only galaxies in the luminosity
range from $M^*+3.5$ to $M^*-2.0$ are used. In practice,  an
initial value of $M^*$ was chosen and an iterative procedure was
used to determine $M^*$ and other LF parameters. Some extreme
bright galaxies are rejected automatically by this selection.

As comparison, we also estimate each LF using the step-wise
maximum-likelihood (SWML) method of Efstathious et al. (1988). Here
the LF is represented by a non-parametric step function and the
maximization of the likelihood function is used to determined the
relative amplitudes at all the steps.

\subsection{Results}
\label{sec:result}

\begin{figure}
  \includegraphics[height=0.5\textwidth,angle=0]{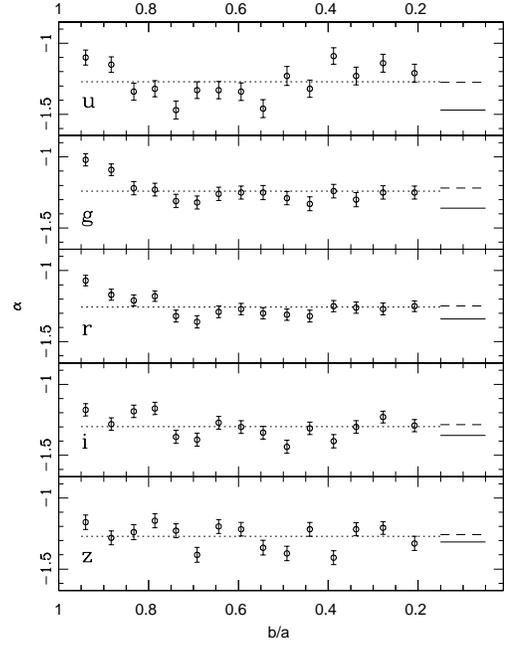}\\
  \caption{The dependence of the faint-end slope of the LF, $\alpha$,
  on the axis ratios of galaxies ($b/a$). Dotted lines for
  $\langle \alpha_{\rm sub}\rangle$ of table~\ref{tab:LF-bin}; Solid lines for $\alpha_{\rm obs}$ of table~\ref{tab:LF-raw};
  Dashed line for $\alpha_{0}$ of table~\ref{tab:LF-corr} (see \S~\ref{sec:model}).}
  \label{fig:alpha15bin}\vspace{8mm}
\end{figure}

\begin{figure}
  \includegraphics[height=0.5\textwidth,angle=0]{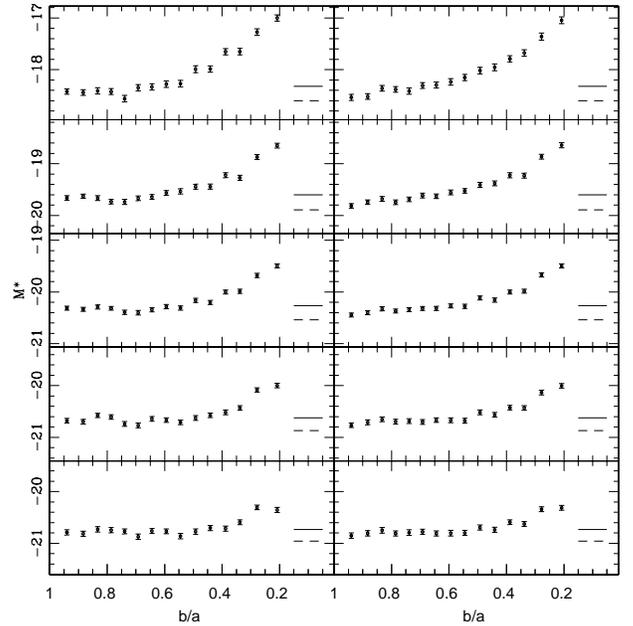}\\
  \caption{The dependence of the characteristic magnitude of the LF,
  $M^*$, on the axis ratios of galaxies ($b/a$).
  In the left panel, the values of $M^*$ are taken from the fit
  with both $M^*$ and $\alpha$ as free parameters,
  In the right panel, the values of $M^*$ are obtained
  from the fit assuming $\alpha=\langle\alpha_{\rm sub}\rangle$.
  Solid lines for $M^*_{\rm obs}$ of  table~\ref{tab:LF-raw}; Dashed line for
   $M_{0}$ of table~\ref{tab:LF-corr} (see \S~\ref{sec:model}).}
  \label{fig:m15bin}\vspace{8mm}
\end{figure}

 In Table~\ref{tab:LF-raw} we list the fitting parameters of the
LFs of spiral galaxies in the 5 SDSS bands, using the STY method.
SWML method gives similar results; they are not presented here in
order to save space. The subscript 'obs' in $\alpha_{\rm obs}$ and
$M^*_{\rm obs}$ denote that these are derived directly from
observational data, to distinguish them from the quantities for
other LFs  discussed in  this paper.  Comparing the results with
that obtained by Blanton et.al. (2001), for all SDSS galaxies, we
see that our LFs are steeper in the faint end (i.e. $\alpha$ is
more negative) in all 5 bands. This is consistent with the fact
that the faint-end slope for late-type galaxies is steeper than
that for early types (Nakamura et al. 2003). In addition, the
typical color of galaxies in our sample, as represented by the
values of $M_{\rm obs}^*$ in the 5 bands is bluer that for the
total sample, as is expected from the fact that spiral galaxies
are on average bluer than early-type galaxies.
\begin{table}
\centering \caption{Parameters of Luminosity Function for
Spiral Galaxies in Each of the 5 SDSS Bands}
\label{tab:LF-raw}
\begin{tabular}{c c c c c}
\hline\hline
Band & $N_{\rm spiral}$ & $\alpha_{\rm obs}$ & $M^*_{\rm obs}$  &\\
\hline
 $u$ & 23757 &  -1.47$\pm$0.02 & -18.32$\pm$0.02  & \\
 $g$ & 42620 &  -1.36$\pm$0.01 & -19.60$\pm$0.01  & \\
 $r$ & 60523 &  -1.34$\pm$0.01 & -20.26$\pm$0.01  & \\
 $i$ & 51593 &  -1.36$\pm$0.01 & -20.62$\pm$0.01  & \\
 $z$ & 43030 &  -1.31$\pm$0.01 & -20.74$\pm$0.01  & \\
\hline
\end{tabular}\vspace{8mm}
\end{table}
 To study the dependence of the LF on galaxy inclinations,
we divide  sample of each band into 15 subsamples according to the
apparent axis ratio $b/a$ in the $r$ band (see
\S~\ref{sec:inclination} and Figure~\ref{fig:cos-theta} for
details). LF is estimated for each of the subsamples in each of
the 5 SDSS wave bands. As an example, Figure~\ref{fig:lf} shows
the LFs for two extreme subsamples, one for nearly edge-on
galaxies, and the other for nearly face-on galaxies, in the
$r$-band. Note that the results obtained from the SWML method
(open and solid circles) match extremely well with those obtained
from the STY method (solid lines), showing that the Schechter
function is a valid assumption for the LFs for spiral galaxies
with different inclinations. Clearly, both subsamples have similar
$\alpha$, but edge-on galaxies are systematically fainter than
face-on galaxies, by about $\sim0.65$ magnitude. In
Table~\ref{tab:LF-bin}, we list the LF parameters obtained from
the SYT method for all the inclination subsamples, and we use the
subscript 'sub' to denote the parameters for subsamples. The
changes of $\alpha_{\rm sub}$ and $M^*_{\rm sub}$ with $b/a$ are
also plotted in Figure~\ref{fig:alpha15bin} and in the left panel
of Figure~\ref{fig:m15bin}, respectively. There is a systematic
change of $M^*_{\rm sub}$, with face-on galaxies having a brighter
characteristic magnitude than edge-on galaxies. This change of
$M^*_{\rm sub}$ with inclination becomes smoother if we fix the
faint-end slope $\alpha$ to be the average value of the 15
sub-samples (dotted lines in Figure~\ref{fig:alpha15bin}), as one
can see from the right panel of Figure~\ref{fig:m15bin}. These
values are denoted as   $M^*_{\rm sub,2}$ and also listed in
Table~\ref{tab:LF-bin}. In addition, the inclination-dependence of
$M^*_{\rm sub}$ is stronger in blue band, as is expected if the
dependence is caused by dust extinction.

A systematic change of $\alpha$ with $b/a$ is expected if dust
extinction depends on galaxy luminosity for $L< L^*$. The absence
of any strong systematic dependence of $\alpha$ on $b/a$ in the
data suggests that dust extinction does not depend on galaxy
luminosity strongly, at least for sub-$L^*$ galaxies. Note,
however, that the dust extinction considered here is relative to
that for face-on galaxies. If dust extinction in face-on galaxies
depends strongly on luminosity, then a strong luminosity-dependent
dust extinction cannot be excluded. We will discuss this in more
detail in \S~\ref{sec:LDE}.

\thispagestyle{empty} \setlength{\voffset}{-25mm}
\begin{table}\begin{center}
\caption{Luminosity Function Parameters for Spiral Galaxies with
Different Inclinations. }\label{tab:LF-bin} {\tiny
\begin{tabular}{ c c c c c c c c c c }
\hline\hline
\\
\small Band & \small $N_{\rm sub,1}$ & \small $(b/a)_{\rm med}$ &
\small$(\cos\theta)_{\rm med}$ & \small$\alpha_{\rm sub}$ &
\small$M^*_{\rm sub,1}$
& \small $N_{\rm sub,2}$ & \small $\langle\alpha_{\rm sub}\rangle$ & \small $M^*_{\rm sub,2}$\\
\\
\hline
 \small $u$& 1256 & 0.21 & 0.09 & -1.21$\pm$0.09 & -17.00$\pm$0.06 & 1260 & -1.27 & -17.04$\pm$0.06\\
     & 1216 & 0.28 & 0.17 & -1.14$\pm$0.08 & -17.27$\pm$0.06 & 1223 & . & -17.36$\pm$0.07\\
     & 1373 & 0.34 & 0.24 & -1.23$\pm$0.08 & -17.65$\pm$0.06 & 1374 & . & -17.68$\pm$0.06\\
     & 1440 & 0.39 & 0.29 & -1.09$\pm$0.07 & -17.65$\pm$0.06 & 1451 & . & -17.79$\pm$0.06\\
     & 1519 & 0.44 & 0.36 & -1.32$\pm$0.07 & -17.99$\pm$0.07 & 1516 & . & -17.95$\pm$0.06\\
     & 1544 & 0.49 & 0.42 & -1.23$\pm$0.07 & -17.99$\pm$0.06 & 1545 & . & -18.02$\pm$0.06\\
     & 1658 & 0.55 & 0.49 & -1.46$\pm$0.07 & -18.27$\pm$0.07 & 1657 & . & -18.15$\pm$0.06\\
     & 1604 & 0.59 & 0.54 & -1.34$\pm$0.07 & -18.28$\pm$0.06 & 1606 & . & -18.24$\pm$0.07\\
     & 1672 & 0.64 & 0.60 & -1.33$\pm$0.07 & -18.33$\pm$0.06 & 1672 & . & -18.30$\pm$0.06\\
     & 1685 & 0.69 & 0.66 & -1.33$\pm$0.07 & -18.35$\pm$0.06 & 1685 & . & -18.31$\pm$0.05\\
     & 1665 & 0.74 & 0.72 & -1.47$\pm$0.07 & -18.56$\pm$0.07 & 1664 & . & -18.41$\pm$0.07\\
     & 1750 & 0.79 & 0.77 & -1.32$\pm$0.07 & -18.42$\pm$0.06 & 1746 & . & -18.38$\pm$0.05\\
     & 1709 & 0.83 & 0.82 & -1.34$\pm$0.07 & -18.41$\pm$0.06 & 1705 & . & -18.36$\pm$0.06\\
     & 1765 & 0.88 & 0.88 & -1.15$\pm$0.06 & -18.44$\pm$0.06 & 1769 & . & -18.52$\pm$0.06\\
     & 1860 & 0.94 & 0.94 & -1.10$\pm$0.06 & -18.42$\pm$0.05 & 1870 & . & -18.54$\pm$0.05\\
\hline
 \small $g$ & 2672 & 0.21 & 0.09 & -1.25$\pm$0.05 & -18.65$\pm$0.05 & 2671 & -1.24 & -18.64$\pm$0.05\\
     & 2598 & 0.28 & 0.17 & -1.25$\pm$0.05 & -18.87$\pm$0.05 & 2598 & . & -18.86$\pm$0.05\\
     & 2726 & 0.34 & 0.24 & -1.30$\pm$0.05 & -19.27$\pm$0.05 & 2726 & . & -19.23$\pm$0.05\\
     & 2788 & 0.39 & 0.29 & -1.24$\pm$0.05 & -19.22$\pm$0.04 & 2788 & . & -19.22$\pm$0.05\\
     & 2841 & 0.44 & 0.36 & -1.33$\pm$0.05 & -19.44$\pm$0.05 & 2842 & . & -19.38$\pm$0.04\\
     & 2834 & 0.49 & 0.42 & -1.29$\pm$0.05 & -19.44$\pm$0.05 & 2834 & . & -19.41$\pm$0.05\\
     & 2894 & 0.55 & 0.49 & -1.25$\pm$0.05 & -19.53$\pm$0.05 & 2897 & . & -19.52$\pm$0.04\\
     & 2853 & 0.59 & 0.54 & -1.25$\pm$0.05 & -19.56$\pm$0.05 & 2852 & . & -19.56$\pm$0.05\\
     & 2887 & 0.64 & 0.60 & -1.26$\pm$0.05 & -19.65$\pm$0.05 & 2889 & . & -19.63$\pm$0.04\\
     & 2921 & 0.69 & 0.66 & -1.32$\pm$0.05 & -19.67$\pm$0.05 & 2920 & . & -19.61$\pm$0.05\\
     & 2897 & 0.74 & 0.72 & -1.31$\pm$0.05 & -19.74$\pm$0.05 & 2896 & . & -19.69$\pm$0.04\\
     & 2925 & 0.79 & 0.77 & -1.23$\pm$0.05 & -19.74$\pm$0.05 & 2926 & . & -19.75$\pm$0.05\\
     & 2839 & 0.83 & 0.82 & -1.22$\pm$0.05 & -19.66$\pm$0.05 & 2838 & . & -19.68$\pm$0.04\\
     & 3008 & 0.88 & 0.88 & -1.09$\pm$0.05 & -19.63$\pm$0.04 & 3014 & . & -19.74$\pm$0.04\\
     & 2981 & 0.94 & 0.94 & -1.02$\pm$0.05 & -19.66$\pm$0.04 & 2983 & . & -19.82$\pm$0.04\\
\hline
 \small $r$ & 4093 & 0.21 & 0.09 & -1.25$\pm$0.04 & -19.50$\pm$0.04 & 4093 & -1.25 & -19.50$\pm$0.04\\
     & 4075 & 0.28 & 0.17 & -1.27$\pm$0.04 & -19.68$\pm$0.04 & 4075 & . & -19.67$\pm$0.04\\
     & 4047 & 0.34 & 0.24 & -1.26$\pm$0.04 & -19.99$\pm$0.04 & 4048 & . & -19.98$\pm$0.04\\
     & 4026 & 0.39 & 0.29 & -1.25$\pm$0.04 & -20.00$\pm$0.04 & 4027 & . & -20.00$\pm$0.04\\
     & 4011 & 0.44 & 0.36 & -1.32$\pm$0.04 & -20.20$\pm$0.04 & 4010 & . & -20.16$\pm$0.04\\
     & 4092 & 0.49 & 0.42 & -1.31$\pm$0.04 & -20.16$\pm$0.04 & 4089 & . & -20.11$\pm$0.04\\
     & 3987 & 0.55 & 0.49 & -1.30$\pm$0.04 & -20.31$\pm$0.04 & 3988 & . & -20.28$\pm$0.04\\
     & 3987 & 0.59 & 0.54 & -1.27$\pm$0.04 & -20.28$\pm$0.04 & 3987 & . & -20.27$\pm$0.04\\
     & 4072 & 0.64 & 0.60 & -1.29$\pm$0.04 & -20.34$\pm$0.04 & 4071 & . & -20.32$\pm$0.04\\
     & 4042 & 0.69 & 0.66 & -1.36$\pm$0.04 & -20.40$\pm$0.04 & 4043 & . & -20.32$\pm$0.04\\
     & 4030 & 0.74 & 0.72 & -1.32$\pm$0.04 & -20.39$\pm$0.04 & 4033 & . & -20.34$\pm$0.04\\
     & 4061 & 0.79 & 0.77 & -1.18$\pm$0.04 & -20.32$\pm$0.04 & 4059 & . & -20.37$\pm$0.04\\
     & 3979 & 0.83 & 0.82 & -1.21$\pm$0.04 & -20.29$\pm$0.04 & 3983 & . & -20.33$\pm$0.04\\
     & 4081 & 0.88 & 0.88 & -1.17$\pm$0.04 & -20.34$\pm$0.04 & 4085 & . & -20.40$\pm$0.04\\
     & 4014 & 0.94 & 0.94 & -1.07$\pm$0.04 & -20.31$\pm$0.04 & 4016 & . & -20.44$\pm$0.04\\
\hline
 \small $i$ & 3700 & 0.21 & 0.09 & -1.29$\pm$0.04 & -20.00$\pm$0.04 & 3700 & -1.30 & -20.00$\pm$0.04\\
     & 3603 & 0.28 & 0.17 & -1.23$\pm$0.05 & -20.08$\pm$0.04 & 3604 & . & -20.13$\pm$0.04\\
     & 3585 & 0.34 & 0.24 & -1.30$\pm$0.05 & -20.43$\pm$0.04 & 3585 & . & -20.43$\pm$0.05\\
     & 3543 & 0.39 & 0.29 & -1.40$\pm$0.05 & -20.52$\pm$0.05 & 3546 & . & -20.43$\pm$0.04\\
     & 3456 & 0.44 & 0.36 & -1.31$\pm$0.05 & -20.57$\pm$0.05 & 3456 & . & -20.56$\pm$0.04\\
     & 3503 & 0.49 & 0.42 & -1.44$\pm$0.05 & -20.62$\pm$0.05 & 3500 & . & -20.52$\pm$0.05\\
     & 3394 & 0.55 & 0.49 & -1.34$\pm$0.05 & -20.71$\pm$0.04 & 3393 & . & -20.68$\pm$0.05\\
     & 3364 & 0.59 & 0.54 & -1.30$\pm$0.05 & -20.67$\pm$0.05 & 3365 & . & -20.67$\pm$0.05\\
     & 3401 & 0.64 & 0.60 & -1.27$\pm$0.05 & -20.64$\pm$0.04 & 3402 & . & -20.67$\pm$0.04\\
     & 3332 & 0.69 & 0.66 & -1.39$\pm$0.05 & -20.77$\pm$0.05 & 3336 & . & -20.70$\pm$0.05\\
     & 3355 & 0.74 & 0.72 & -1.37$\pm$0.05 & -20.74$\pm$0.05 & 3355 & . & -20.69$\pm$0.05\\
     & 3385 & 0.79 & 0.77 & -1.17$\pm$0.05 & -20.61$\pm$0.04 & 3387 & . & -20.70$\pm$0.04\\
     & 3270 & 0.83 & 0.82 & -1.19$\pm$0.05 & -20.58$\pm$0.04 & 3269 & . & -20.65$\pm$0.04\\
     & 3381 & 0.88 & 0.88 & -1.28$\pm$0.05 & -20.70$\pm$0.05 & 3379 & . & -20.71$\pm$0.05\\
     & 3338 & 0.94 & 0.94 & -1.18$\pm$0.05 & -20.68$\pm$0.04 & 3335 & . & -20.77$\pm$0.04\\
\hline
\small $z$  & 3275 & 0.21 & 0.09 & -1.32$\pm$0.05 & -20.35$\pm$0.05 & 3273 & -1.27 & -20.31$\pm$0.05\\
     & 3229 & 0.28 & 0.17 & -1.21$\pm$0.05 & -20.31$\pm$0.05 & 3233 & . & -20.34$\pm$0.04\\
     & 3147 & 0.34 & 0.24 & -1.22$\pm$0.05 & -20.59$\pm$0.05 & 3147 & . & -20.63$\pm$0.05\\
     & 3086 & 0.39 & 0.29 & -1.42$\pm$0.05 & -20.72$\pm$0.05 & 3076 & . & -20.59$\pm$0.05\\
     & 2928 & 0.44 & 0.36 & -1.22$\pm$0.06 & -20.70$\pm$0.05 & 2929 & . & -20.74$\pm$0.05\\
     & 2934 & 0.49 & 0.42 & -1.39$\pm$0.06 & -20.77$\pm$0.05 & 2932 & . & -20.69$\pm$0.05\\
     & 2859 & 0.55 & 0.49 & -1.35$\pm$0.06 & -20.86$\pm$0.05 & 2857 & . & -20.80$\pm$0.05\\
     & 2724 & 0.59 & 0.54 & -1.22$\pm$0.06 & -20.77$\pm$0.05 & 2724 & . & -20.80$\pm$0.05\\
     & 2775 & 0.64 & 0.60 & -1.20$\pm$0.06 & -20.76$\pm$0.05 & 2778 & . & -20.81$\pm$0.05\\
     & 2755 & 0.69 & 0.66 & -1.40$\pm$0.06 & -20.87$\pm$0.05 & 2752 & . & -20.78$\pm$0.05\\
     & 2640 & 0.74 & 0.72 & -1.23$\pm$0.06 & -20.77$\pm$0.05 & 2641 & . & -20.80$\pm$0.05\\
     & 2704 & 0.79 & 0.77 & -1.16$\pm$0.06 & -20.75$\pm$0.05 & 2706 & . & -20.81$\pm$0.05\\
     & 2603 & 0.83 & 0.82 & -1.24$\pm$0.06 & -20.73$\pm$0.05 & 2604 & . & -20.75$\pm$0.05\\
     & 2682 & 0.88 & 0.88 & -1.28$\pm$0.06 & -20.81$\pm$0.05 & 2684 & . & -20.81$\pm$0.05\\
     & 2705 & 0.94 & 0.94 & -1.17$\pm$0.06 & -20.79$\pm$0.05 & 2706 & . & -20.85$\pm$0.05\\
\hline
\end{tabular}
}
\end{center}
\tablenotetext{}{Samples of all 5 bands are divided into 15
sub-samples by $b/a$, while $(b/a)_{\rm med}$ and
$(\cos\theta)_{\rm med}$ show the median values of inclination of
each subsample. ($\alpha_{\rm sub}, M^*_{\rm sub,1}$) denoted the
fitting parameter of LF for each sub-sample, and $M^*_{\rm sub,2}$
are fitting values of characteristic magnitude when we fixed the
slope parameter to be the average value of 15 sub-samples,
$\langle\alpha_{\rm sub}\rangle$. \vspace{8mm}}
\end{table}
\clearpage
\clearpage

\section{Internal dust extinction}\label{sec:IDE}

  Having shown that the LF of spiral galaxies depends systematically
on the inclination, we now model such dependence in terms of dust
extinction.

\subsection{Empirical extinction models}
\label{sec:model}

 As a simple model for the dust extinction, we assume that
the change in magnitude of a galaxy, $\Delta M$, due to dust
extinction is proportional to $\log(b/a)$, so that
\begin{equation}\label{eq:curves1}
\Delta M_1(b/a)=M(b/a)-M_1(1)=-\gamma_1\log(b/a)\,,
\end{equation}
where $\gamma_1$ parameterizes the amplitude of the dust extinction
relative to the face-on value (e.g. Giovanelli et al. 1994;
Masters, Giovanelli \& Haynes 2003). As discussed in
\S~\ref{sec:inclination}, a quantity that may better describe the
inclination is the {\it corrected} cosine of the inclination
angle, $\cos\theta$. We therefore consider another model in which
\begin{equation}\label{eq:curves2}
\Delta M_2(\cos\theta)= M(\cos\theta)-M_2(1)=-\gamma_2\log(\cos\theta)\,.
\end{equation}
In the above expressions, the subscripts, `1' and `2' denote
quantities in the two models. Note that $M_1(1)$ and $M_2(1)$
are the magnitudes corresponding to $b/a=1$ and $\cos\theta=1$,
i.e. face-on disk.

\begin{figure}
  \includegraphics[height=0.5\textwidth,angle=0]{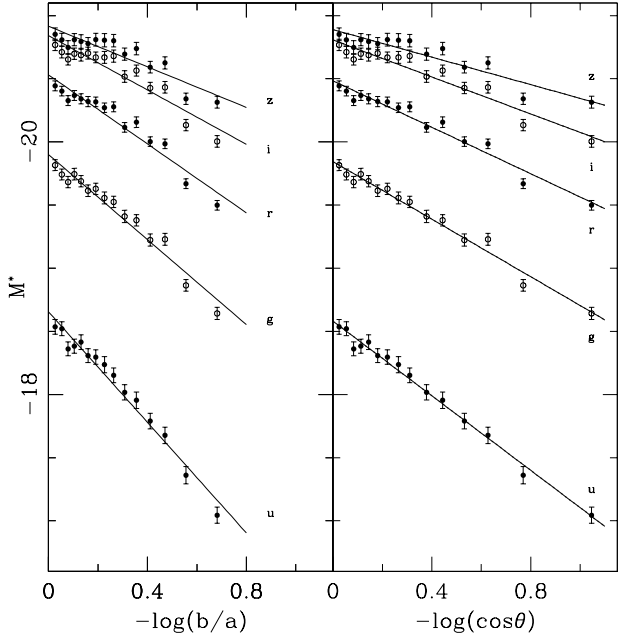}\\
  \caption{The inclination-dependence of $M^*$ ($M^*_{\rm sub,2}$
  listed in table~\ref{tab:LF-bin}, see text for details) and its best linear fit.
   Solid and open circles are used to distinguish
   two successive wave-bands.}
  \label{fig:gamma-fit1}\vspace{8mm}
\end{figure}

  Since the inclination-dependence of $\alpha_{\rm sub}$ is not strong,
the main effect of dust extinction is to change $M^*_{\rm sub}$.
The change of $M^*_{\rm sub}$ with inclination may be used
to infer the average dust extinction for all spiral galaxies in
the sample. The results of the least square fits of equations
(\ref{eq:curves1}) and (\ref{eq:curves2}) to the observed
$M^*_{\rm sub,2}$-$\log(b/a)$ and $M^*_{\rm
sub,2}$-$\log(\cos\theta)$ relations are shown in
Fig.~\ref{fig:gamma-fit1}, while the values of the best fit
$\gamma$ and $M^*(1)$ are listed in Table~\ref{tab:gama} along
with the corresponding $\chi^2$ values. There is significant
difference between the results based on the apparent axis ratio
$b/a$ and those based on the corrected inclination angles
$\theta$. For all of the 5 wave bands, $\chi^2_2 < \chi^2_1$,
suggesting that a linear model works better for the corrected
inclination angle. The values of $\gamma_2$ are all smaller than
$\gamma_1$, because the relation between $b/a$ and $\cos\theta$
has a slope that is different from 1 (see
Figure~\ref{fig:cos-theta}). However, the ratios between
the $\gamma$ values in different wave bands are quite similar
in the two models. As we will see in \S~\ref{sec:n-fit}, it
is these ratios that describe the dust extinction curve.
\begin{table}\caption{Fitting parameters of the linear relationship
between $M^*$ and the logarithmic of the inclination. Also see
solid lines in Figure~\ref{fig:gamma-fit1}.}\label{tab:gama}
\begin{tabular}{c c c c r c c c r}
\hline\hline
 Band & & $M^*_{1}(1)$ & $\gamma_1$ & $\chi^2_1$ & &
$M^*_{2}(1)$ & $\gamma_2$ & $\chi^2_2$\\
\hline
$u$&  & -18.65$\pm$0.03 & 2.19$\pm$0.08 & 18.82 & & -18.58$\pm$0.02 & 1.48$\pm$0.06 & 7.24 \\
$g$&  & -19.90$\pm$0.02 & 1.68$\pm$0.06 & 31.36 & & -19.84$\pm$0.02 & 1.14$\pm$0.04 & 15.31 \\
$r$&  & -20.53$\pm$0.02 & 1.37$\pm$0.05 & 48.28 & & -20.48$\pm$0.02 & 0.92$\pm$0.04 & 26.29 \\
$i$&  & -20.84$\pm$0.02 & 1.08$\pm$0.06 & 41.61 & & -20.81$\pm$0.02 & 0.73$\pm$0.04 & 26.77 \\
$z$&  & -20.92$\pm$0.02 & 0.80$\pm$0.07 & 30.98 & & -20.89$\pm$0.02 & 0.54$\pm$0.04 & 23.71 \\
\hline
\\
\end{tabular}\vspace{8mm}
\end{table}
In the maximum-likelihood estimate of the LF used here, dust
extinction can be incorporated in a more elegant way. Since
equation (\ref{eq:curves2}) provides a reasonable description of
the mean relation between the dust extinction and the inclination
angle $\theta$, a similar relation may be used for individual
galaxies. In this case, the luminosity function of spiral galaxies
with observed luminosity $L$ and inclination angle $\theta$ can be
parameterized by
\begin{equation}
\label{eq:lf-gamma}
\phi(L,\theta)\propto
\left[(L/L^*) (\cos\theta)^{-0.4\gamma}\right]^\alpha
\exp\left[-(L/L^*)(\cos\theta)^{-0.4\gamma}\right]\,,
\end{equation}
Thus, the value of $\gamma$ (assumed to be independent of galaxy
luminosity) can also be determined through the maximum likelihood
method (either the STY method or the SWML method) that determines
the LF. We have carried out such a maximum likelihood analysis
using the STY method. Again, an iterative procedure is used so
that only galaxies with luminosities in the range between
$M^*+3.5$ and $M^*-2.0$ are used in the fitting. Since
$\cos\theta$ for individual galaxies are obtained statistically
from Monte-Carlo simulations, different realizations may give
different results. We therefore construct 25 different
realizations and obtain the values of $\alpha$, $M^*$ and $\gamma$
for each of them. The average values of $\alpha$, $M^*$ and
$\gamma$ obtained in this way are listed in
Table~\ref{tab:LF-corr} as $\alpha_{0}$, $M^*_{0}$ and $\gamma_3$,
where the subscript `0' is used to denote 'face-on' galaxies, and
the subscript `3' in $\gamma_3$ is used to distinguish it from
$\gamma_1$ and $\gamma_2$ obtained from the $M^*_{\rm
sub,2}$-$\log(b/a)$ relation and the $M^*_{\rm
sub,2}$-$\log(\cos\theta)$ relation. The uncertainties on these
parameters include both fitting errors and the scatter among the
25 realizations. The error bars are all very small, because each
sample is sufficiently large to represent the distribution of
$\cos\theta$ faithfully. Note that the values of $\gamma_3$ are
similar to that of $\gamma_2$, indicating that the method based on
$M^*$ and  that on individual galaxies give the same results.
\begin{table}
\centering \caption{Corrected Luminosity Function
(eq.~\ref{eq:lf-gamma}) for Spiral Galaxies in Each of the SDSS
Band} \label{tab:LF-corr}
\begin{tabular}{c c c c c}
\hline\hline
Band & $N_{\rm spiral}$ & $\alpha_{0}$ & $M^*_{0}$ & $\gamma_3$ \\
\hline
$u$  & 23380 &  -1.28$\pm$0.02 & -18.60$\pm$0.02 &   1.59$\pm$0.05 \\
$g$  & 42150 &  -1.22$\pm$0.01 & -19.89$\pm$0.01 &   1.24$\pm$0.03 \\
$r$  & 60100 &  -1.25$\pm$0.01 & -20.54$\pm$0.01 &   0.92$\pm$0.02 \\
$i$  & 51290 &  -1.28$\pm$0.01 & -20.87$\pm$0.01 &   0.80$\pm$0.02 \\
$z$  & 42820 &  -1.26$\pm$0.01 & -20.96$\pm$0.01 &   0.65$\pm$0.03 \\
 \hline
\end{tabular}
\tablenotetext{}{. These values represent the parameters for pure
face-on spiral galaxies.\vspace{8mm}}
\end{table}

\subsection{Extinction curves}\label{sec:n-fit}

 The wavelength-dependence of $\gamma$ can be used to constrain
the extinction curve for spiral galaxies. As a simple model, we
assume that the effective extinction curve in the optical has a
power-law form,
\begin{equation}\label{eq:tau}
\tau_\lambda(\cos \theta)=\tau_V(\cos \theta)
(\lambda/5500\AA)^{-n}\,,
\end{equation}
where $\lambda$ is the wavelength, $\tau_\lambda$ is the optical depth
at $\lambda$, $\tau_V$ is the optical depth at the $V$-band
(centered at $\lambda=5500{\rm \AA}$) and \emph{is a function of
inclination angle}, and $n$ is the power index describing the shape
of the dust extinction curve. If we neglect the difference in the
dust extinction curve for different galaxies and compare the
dust extinction at given inclination angle but in different
wavelength, then $\gamma$ is proportional to $\tau$. Thus, we
can write
\begin{equation}\label{eq:gamma}
\gamma_\lambda=\gamma_V(\lambda/5500\AA)^{-n}\,,
\end{equation}
where $\gamma_V$ is the value of $\gamma$ at $5500\AA$.

\begin{figure}
  \includegraphics[height=0.45\textwidth,angle=0]{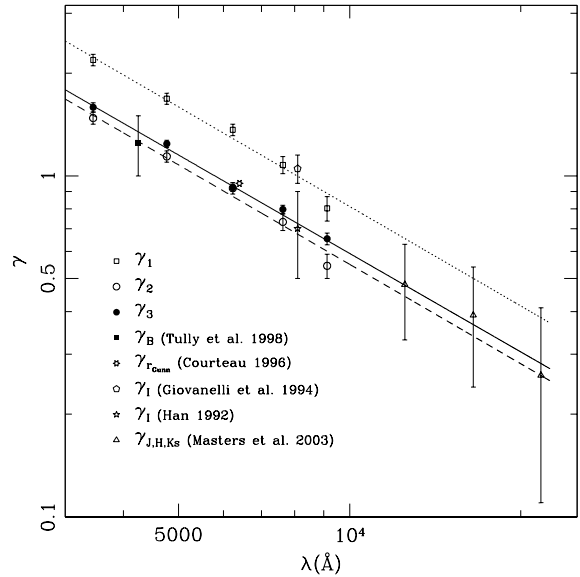}\\
  \caption{The best fit of the extinction curve in optical wavebands.
  The values of $\gamma$ are those given by the different methods
  described in \S~\ref{sec:n-fit}.
  Data points show the values of $\gamma$ obtained
  by other investigators.}
  \label{fig:n-fit1}\vspace{8mm}
\end{figure}

We use a least-square fitting to estimate $\gamma_V$ and $n$ from
the values of $\gamma$ in the 5 SDSS bands, and the results are
listed in Table~\ref{tab:n} for the three cases, Cases 1, 2 and 3,
corresponding to the use of $\gamma_1$, $\gamma_2$ and $\gamma_3$,
respectively. In Figure~\ref{fig:n-fit1}, we plot the fitting
results along with the data of $\gamma$ used in the fit. As one
can see, the power-law model is a good assumption in all cases. It
is also interesting to note that, although the values of
$\gamma_1$, $\gamma_2$ and $\gamma_3$ are different, the power-law
indices  obtained from them are very similar,
with $n=0.96\pm0.04$.
\begin{table}
  \caption{Fitting Results for the Extinction Curve (eq.~\ref{eq:gamma}).}\label{tab:n}
  \begin{center}
\begin{tabular}{c c c c }
\hline\hline
  & $\gamma_V$ & $n$ & $\chi^2$ \\
\hline
Case~1 (using $\gamma_1$)    & 1.45$\pm$0.02 & 0.97$\pm$0.07 & 3.94\\
Case~2 (using $\gamma_2$)    & 0.98$\pm$0.02 & 0.97$\pm$0.07 & 3.71\\
Case~3 (using $\gamma_3$)    & 1.05$\pm$0.01 & 0.96$\pm$0.04 & 1.50\\
 \hline
\end{tabular}
\end{center}\vspace{8mm}
\end{table}
(\ref{eq:tau}) we obtained from the SDSS data are extrapolated to
the near-infrared, and compared with the data obtained by Masters
et al. (2003) for spiral galaxies with $\log (a/b)<0.5$ in the
2MASS: $\gamma_J=0.48\pm0.15$, $\gamma_H=0.39\pm0.15$ and
$\gamma_{K_S}=0.26\pm0.15$. In addition, we also plot results
obtained by others in various bands. Tully et al. (1998) gave
$\gamma_B= 1.0$ -1.5 for different types of spiral galaxies.
Courteau (1996) obtained $\gamma_{r_{\rm Gunn}}= 0.95$ using 349
Sb-Sc UGC galaxies with $0.27 < \log(a/b) < 0.70$. Han (1992)
found $\gamma_I$ to be in the range from $0.51$ to $0.91$ for
different types of galaxies in the $I$ band. Finally, Giovanelli
et al. (1994) obtained $\gamma_I=0.95$ - $1.15$. Remarkably, all
of the results follow well the power-law we obtained from  the
SDSS data.

It should be pointed out, however, that the dust extinction we
obtained here is the effective extinction for an entire galaxy,
rather than that from some specific regions within a galaxy. Such
extinction depends not only on the properties of the dust grains
but also on the distribution of dust in individual galaxies. The
index, $n=0.96\pm0.04$, obtained is shallower than that obtained
by observations of the Milky Way, the LMC and the SMC,  $n=1.1$ -
$1.5$, but steeper than the value $n=0.7$ obtained by Charlot et
al. (2000) based on the assumption that dust in a spiral galaxy
has a patchy distribution. Our result of $n\sim1$ implies that
dust distribution in spiral galaxies on average are not as patchy
as assumed in Charlot et al. (2000).

\section{Extinction-corrected luminosity function}
\label{sec:discussion}

\subsection{Luminosity-dependence of extinction}
\label{sec:LDE}

 In the analysis presented above, we have assumed that
the value of $\gamma$ is independent of luminosity. In this
subsection we examine whether or not this assumption is valid.
To do this, we consider a model in which $\gamma$ is
a linear function of the magnitude,
\begin{equation}\label{eq:beta}
    \gamma (M)=\gamma^*[1+\beta(M-M^*)]\,.
\end{equation}
We treat $\beta$ as the fourth free parameter in
fitting the LF. The result shows that $\beta$ has a very small value
so that the difference between $\gamma$ and $\gamma^*$
is less than 5\%. Furthermore, if $\alpha$ is kept
constant in the fitting, $\beta$ is even smaller. Thus,
the luminosity-dependence of $\gamma$ is weak and will
be neglected in the following. Note, however, that $\gamma$ in
our definition only
characterizes the extinction at a given inclination relative to
that of face-on galaxies. Thus, even if $\gamma$ is independent of
galaxy luminosity, the absolute value of the extinction may be
luminosity-dependent, if the extinction of face-on galaxies
depends on luminosity. Unfortunately our method, which is based on
a comparison between inclined galaxies with face-on galaxies,
cannot be used to obtained the absolute value of the extinction.
Because of this, here we use an independent set of observational
data to probe the luminosity-dependence of extinction and to
examine how such data can be used together with our results to
obtain an extinction-corrected luminosity function for spiral
galaxies.

Kauffmann et al. (2003) estimated the dust extinction in the SDSS
$z$-band, $A_z$, for galaxies in the SDSS DR2, using the $4000\AA$
break strength and the Balmer absorption line index $H\delta_A$ to
constrain the star formation history and then estimating the dust
extinction from the difference between the model and observed
$g-r$ and $r-i$ colors. We combine their data with our sample. It
should be noted that the $4000{\rm\AA}$ breaks and the ${\rm
H\delta_A}$ indices are estimated from the SDSS fiber spectra
within a 3-arcsec aperture. Thus, the values of $A_z$ given by
Kauffmann et al. may be biased towards the central parts of
galaxies, especially for nearby galaxies with large apparent
sizes. Figure~\ref{fig:Az-Mz} shows $A_z$ versus luminosity for
galaxies in our sample. The small dots in the upper panel show all
galaxies in the `face-on' subsample with $0.86<b/a<0.91$ . Open
circles with error bars represent the median and their errors
within given bins of $M_z$. As a comparison, the solid circles
show the median values of $A_z$ for `edge-on' galaxies with
$0.25<b/a<0.31$. The solid circles are shifted by 0.5 magnitude
towards the bright end according to the difference in $M^*_{\rm
sub}$ for these two subsamples. It is clear that dust extinction
does depend on luminosity for both face-on and edge-on galaxies.
The dependence is not a simple monotonic relation. The extinction
increases with luminosity at the faint part and decreases slightly
towards the bright end. The behavior at the faint end may indicate
that the dust opacity is larger in brighter galaxies. The slight
decrease of $A_z$ with luminosity at the bright end may be due to
the fact that the contribution of the central bulge, which may be
less dusty, becomes more important for brighter galaxies. Note
that the mean $A_z$-$M_z$ relation for the face-on subsample is
roughly parallel to that for the edge-on subsample, except perhaps
for faint galaxies (see the lower panel of
Figure~\ref{fig:Az-Mz}). Since $\gamma_z \propto \Delta A_z$, this
result implies that $\gamma_z$ is quite independent of luminosity,
consistent with our result based on the luminosity functions.

The difference in $A_z$ between face-on and edge-on galaxies is
about 0.4 magnitude, in good agreement with the difference in
$M_{\rm sub}^*$ (~0.5 mag) between these two subsamples.
We have also checked the $M_z$-dependence of $A_z$
for other subsamples of $b/a$, and found that the relations
are all roughly parallel to each other, with the relative
amplitudes very similar to those obtained from fitting the LFs
(see Table~\ref{tab:LF-bin}). This agreement is remarkable,
because the $A_z$ values given by Kauffmann et al. (2003) are
based on totally different considerations.

\subsection{Extinction-corrected luminosity function}

\subsubsection{Luminosity function of face-on galaxies}

Based on the results presented above, we can have several
estimates for the LF of face-on spiral galaxies. The first one is
based on $(\alpha_{\rm sub}, M_{\rm sub,1}^*)$, or,
$(\langle\alpha_{\rm sub}\rangle, M_{\rm sub,2}^*)$, obtained from
the subsamples with big values of axis ratio (see lines 14 and 15
in Table~\ref{tab:LF-bin}). The second is given by
$(\langle\alpha_{\rm sub}\rangle, M_1^*(1))$ or
$(\langle\alpha_{\rm sub}\rangle, M_2^*(1))$, where $M_1^*(1)$ and
$M_2^*(1)$ are the values of $M^*$ for face-on galaxies obtained
from fitting the $M^*_{sub}$-inclination relations (see
Table~\ref{tab:gama}). The most rigorous estimate may be given by
the values of $( \alpha_{0}, M^*_{0})$ in Table~\ref{tab:LF-corr},
obtained through the maximum likelihood analysis that includes
$\gamma$ as the 3rd parameter (eq.~\ref{eq:lf-gamma}). Note that
for all the 5 SDSS bands, the values of $\langle\alpha_{\rm
sub}\rangle$ and $\alpha_{0}$ are similar, but both are slightly
less negative than $\alpha_{\rm obs}$, the faint end slope of
directly measured LF of spirals (\S~\ref{sec:result}).
Furthermore, unlike $\alpha_{\rm obs}$, which has more negative
values in the bluer bands, both $\langle\alpha_{\rm sub}\rangle$
and $\alpha_{0}$ are quite independent of wave-bands. This may be
explained by the fact that dust extinction not only reduces $L^*$,
but also steepens the LF for spiral galaxies. We have tested this
effect using a Monte Carlo simulation. We generated a sample of
galaxies with a given LF, and with each galaxy assigned a random
orientation. We then made each galaxy dimmer according to its
inclination angle. We found that the resulting LF is steeper than the
original LF.

\begin{figure}
  \includegraphics[height=0.5\textwidth,angle=0]{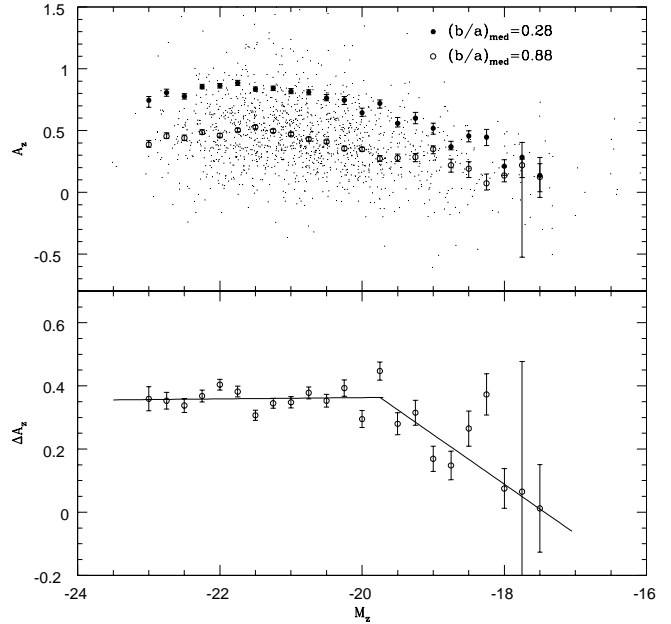}\\
  \caption{The luminosity-dependence of the dust extinction in
   the $z$-band ($A_z$). In the upper panel, small dots are
   the values of $A_z$ obtained by Kauffmann et. al. (2003)
   for individual face-on spiral galaxies with $0.86<b/a<0.91$.
   Open circles with error bars represent the median value of $A_z$
   for  a given $M_z$. Solid circles are the same as
   open ones but for galaxies with $0.25<b/a<0.31$ (almost edge-on).
   The lower panel shows the difference of $A_z$ between face-on
   and edge-on  galaxies. Solid lines here show a
   fitting by eye of the data points.}
  \label{fig:Az-Mz}\vspace{8mm}
\end{figure}

The corrections in the characteristic magnitudes are also
significant. For face-on galaxies, $M_1^*(1)$ , $M_2^*(1)$ and
$M_{0}^*$ are all brighter than $M_{\rm obs}^*$  for about
$0.2$-$0.3$ magnitudes. Figure~\ref{fig:mstar} shows a comparison
of $M^*$s for all the 5 bands. According to the extinction curve
($\tau\propto\lambda^{-1}$), the change in $M^*$ is expected
to be larger for bluer band. That such dependence is not
seen clearly in Figure~\ref{fig:mstar} is largely caused
by the change in $\alpha$ and the degeneracy
between $\alpha$ and $M^*$.

\subsubsection{Intrinsic Luminosity Function}

According to Figure~\ref{fig:Az-Mz}, the values of $A_z$ for
face-on galaxies are typically 0.3-0.4 magnitudes. In principle,
we can use the values of $A_z$ for individual galaxies and
estimate a dust-corrected LF. The LFs in other bands can also be
obtained using the the dust extinction curve. Unfortunately, this
approach is impractical, because the uncertainty in $A_z$ is quite
large ($\Delta A_z \geq 0.2$) and can significantly broaden the
LF. On the other hand, for galaxies with a given luminosity, the
true values of $A_z$ may have intrinsic scatter, which should be
taken into account in the dust correction. Because of this
uncertainty, here we consider two extreme cases. In one, we use
the $A_z$ values given by Kauffmann et al. (2003) to make
correction for each galaxy, and estimate the LF for the
`corrected' sample. The LF parameters obtained in this way are
$\alpha= -1.43 \pm 0.01$ and $M^*= -21.52 \pm 0.01$. In the second
case, we  average the values of $A_z$ for  galaxies with similar
inclination angles, and use them to correct  individual galaxies.
The LF parameters obtained in this way are $\alpha= -1.28 \pm
0.01$ and $M^*= -21.27 \pm 0.01$. As discussed above, the true LF
is expected to be between these two. Our Monte Carlo simulation
showed that the uncertainty in $A_z$, about $0.2$, can account for
most of the difference between these two results. So the second
estimation of LF may close to the real case.

For the other 4 bands, direct estimates of the dust extinction are
not available. We therefore use an assumption of the dust
extinction curve together with the average values of $A_z$ (as a
function of $b/a$) to make dust corrections for individual
galaxies. We list the parameters for the LFs obtained for the
corrected samples in Table~\ref{tab:LF-int} as $\alpha_{\rm int}$
and $M^*_{\rm int}$. For comparison, results are given for two
extinction curves, one with $n=1.0$, as is obtained in this paper,
and the other is $n=0.7$, as given in Charlot et. al. (2000).

\begin{figure}
  \includegraphics[height=0.4\textwidth,angle=-90]{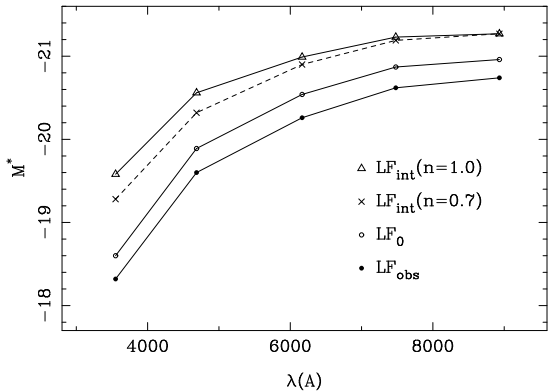}\\
  \caption{A comparison between the values of $M^*$ obtained
   from different cases in different bands.  LF$_{\rm obs}$ labels
   the results of the observational sample without any
   dust correction. LF$_{0}$ labels the results for face-on
   galaxies obtained by correcting the inclination effects.
   LF$_{\rm int}$ labels the results for `extinction-corrected'
   luminosity functions.}
  \label{fig:mstar}\vspace{8mm}
\end{figure}

In Figure~\ref{fig:mstar}, we plot $M^*$ in all the five bands for
different cases: the directly observed LF, the LF for face-on
galaxies, and the intrinsic LFs obtained above. It is clear that
the differences between different cases are quite large. In
particular, $M^*$ for the corrected LF could be $0.5$-$1.2$
magnitudes brighter than the uncorrected one from $z$ to $u$
bands. This suggests that, without correcting internal extinction
of galaxies, the $L^*$ for spiral galaxies can be seriously
under-estimated.
\begin{table}
\centering \caption{Parameters of Luminosity Function for Spiral
Galaxies with correction of dust extinction. } \label{tab:LF-int}
\begin{tabular}{c c c c | c c c}
\hline\hline
 & & n=1.0 & & & n=0.7 &\\
\hline
Band & $N_{\rm spiral}$ & $\alpha_{\rm int}$ & $M^*_{\rm int}$  &$N_{\rm spiral}$ & $\alpha_{\rm int}$ & $M^*_{\rm int}$  \\
\hline
 $u$ & 22887 & -1.46$\pm$0.01 & -19.58$\pm$0.02 & 22902 & -1.50$\pm$0.01 & -19.28$\pm$0.02\\
 $g$ & 41587 & -1.32$\pm$0.01 & -20.56$\pm$0.01 & 41582 & -1.32$\pm$0.01 & -20.32$\pm$0.01\\
 $r$ & 59545 & -1.26$\pm$0.01 & -20.99$\pm$0.01 & 59536 & -1.26$\pm$0.01 &
 -20.90$\pm$0.01\\
 $i$ & 50718 & -1.30$\pm$0.01 & -21.23$\pm$0.01 & 50719 & -1.30$\pm$0.01 &
 -21.19$\pm$0.01\\
 $z$ & 42311 & -1.28$\pm$0.01 & -21.27$\pm$0.01 & 42311 & -1.28$\pm$0.01 &
 -21.27$\pm$0.01\\
\hline
\end{tabular}\vspace{8mm}
\end{table}

\section{Summary and Discussion}
\label{sec:summary}

Using a samples of 61506 spiral galaxies selected  from the SDSS
DR2, we study the luminosity functions of spiral galaxies with
different inclination angles. The apparent axis ratio, $b/a$, is
used as an observational inclination indicator to define
subsamples of spiral galaxies at different inclinations, and we
use a Monte Carlo process to connect $b/a$ to the corrected
inclination angle, $\theta$, disk thickness, $\nu$, and
ellipticity $\epsilon$. There is a systematic change of the LF
with inclination angle: the characteristic luminosity $L^*$
decreases with increasing inclination angle, while the faint-end
slope depends only weakly on inclination.

The inclination-angle dependence of the galaxy luminosity function
is consistent with the expectation of a simple model where the
optical depth is proportional to $\cos\theta$, and we use a
likelihood  method to recover both the amplitude of the extinction
(relative to the face-on value), $\gamma$, and the luminosity
function for galaxies viewed face-on. We found that the results
obtained from different methods are all consistent with each
other, and the characteristic magnitude for face-on spirals is
about $0.2\sim0.3$ magnitudes brighter than the average population.

We found that the value of $\gamma$ is quite independent of
luminosity in a given band. The values of $\gamma$ obtained in this
way for the 5 SDSS bands are used  to constrain the shape of
the extinction curve, assuming $\tau_\lambda = \tau_V
(\lambda/5500\AA)^{-n}$. We find $n=0.96\pm 0.04$.

Using the $z$-band dust extinction given by Kauffmann et
al. (2003), together with the inclination-dependence of the LF we
obtained, we derive an `extinction-corrected' luminosity function
for spiral galaxies. Dust extinction makes a significant change in
$M^*$, and the characteristic luminosity of the `dust-corrected'
LF is about 0.5 to 1.2 magnitude brighter than the uncorrected LF
from the $z$- to the $u$-bands. This suggests that the luminosity
function of spiral galaxies may be significantly underestimated in
blue bands, if internal dust correction is not made. This may have
important implications when comparing model predictions of the
luminosity function with observations.

 As mentioned in the introduction, the dimming in luminosity
from face-on to edge on is expected if galaxy disks are optically
thick. Our results therefore give support to the assumption that
most of the disks in our galaxy sample are optically thick.
Note that the inclination-dependent dimming exists not only for
faint galaxies but for galaxies over the entire luminosity range.
Note also that we are using a well-defined flux-limited sample
and the selection bias is taken into account in the LF estimate.
It is therefore unlikely that the effect we find here is due to
the magnitude limit in the sample. Systematic effects of surface
brightness relative to the inclination is also unlikely to play
an important role here. As discussed in Blanton et al. (2001),
the surface-brightness limit used in the SDSS has negligible
effects on the LF at $M_r<-18$, and so the inclination-dependence
of the LF at $M_r<-18$ cannot be explained as due to the miss of
low-surface face-on galaxies. We therefore conclude that our
results are best explained by assuming that bright disks are
optically thick.

\acknowledgements

The authors thank  Shude Mao,  Jiasheng Huang, Chenggang Shu,
Ruixiang Chang  for their valuable advice and discussion. Thanks
are also due to the anonymous referee for helpful comments. This
research was supported by NSF of China grants No.  10273016,
10333060, and also supported in part by the National Science
Foundation under Grant No. PHY99-07949.

Funding for the creation and distribution of the SDSS Archive has
been provided by the Alfred P. Sloan Foundation, the Participating
Institutions, the National Aeronautics and Space Administration,
the National Science Foundation, the U.S. Department of Energy,
the Japanese Monbukagakusho, and the Max Planck Society. The SDSS
Web site is http://www.sdss.org/.  The SDSS is managed by the
Astrophysical Research Consortium (ARC) for the Participating
Institutions.  The Participating Institutions are The University
of Chicago, Fermilab, the Institute for Advanced Study, the Japan
Participation Group, The Johns Hopkins University, Los Alamos
National Laboratory, the Max-Planck-Institute for Astronomy
(MPIA), the Max-Planck-Institute for Astrophysics (MPA), New
Mexico State University, Princeton University, the United States
Naval Observatory, and the University of
 Washington.



\end{document}